\documentclass[11pt,aps,prd,nofootinbib,superscriptaddress, onecolumn,preprintnumbers,balancelastpage]{revtex4}
\pdfoutput=1
\usepackage{graphicx}
\usepackage{hyperref}
\usepackage{amssymb,amsmath,latexsym,graphics, graphicx,epsfig,multirow,comment,appendix, slashed, verbatim} 
\usepackage{epstopdf}
\usepackage{url}
\usepackage{appendix}
\usepackage{color}

\def\ltap{\ \raise.3ex\hbox{$<$\kern-.75em\lower1ex\hbox{$\sim$}}\ }
\def\gtap{\ \raise.3ex\hbox{$>$\kern-.75em\lower1ex\hbox{$\sim$}}\ }
\newcommand{\gsim}{\lower.7ex\hbox{$\;\stackrel{\textstyle>}{\sim}\;$}}
\newcommand{\lsim}{\lower.7ex\hbox{$\;\stackrel{\textstyle<}{\sim}\;$}}
\def\OO{{\cal O}}

\def\eg{{\it e.g.}}

\newcommand{\Eref}[1]{Eq.~(\ref{#1})}

\newcommand{\GeV}{\,\mathrm{GeV}}

\def\unit{\relax{\rm 1\kern-.26em I}}

\newcommand{\sys}{{\epsilon_{\text{sys}}}}

\newcommand{\be}{\begin{equation}}
\newcommand{\ee}{\end{equation}}
\newcommand{\bea}{\begin{eqnarray}}
\newcommand{\eea}{\end{eqnarray}}

\newcommand{\MET}{$\slashed{E}_T\,\,$}
\newcommand{\go}{\widetilde{g}}

\begin{document}
\begin{flushright}
\mbox{\normalsize SLAC-PUB-15307}
\\
\mbox{\normalsize pi-partphys-309}
\end{flushright}
\vskip 65 pt

\title{
Jet Substructure by Accident
 }

\author{Timothy Cohen}
\affiliation{
SLAC, Stanford University, Menlo Park, CA 94025}

\author{Eder Izaguirre}
\affiliation{Perimeter Institute for Theoretical Physics, Waterloo, ON, N2L 2Y5, Canada}

\author{Mariangela Lisanti}
\affiliation{
PCTS, Princeton University, Princeton, NJ 08544}

\author{Hou Keong Lou}
\affiliation{
Department of Physics, Princeton University, Princeton, NJ 08540}

\begin{abstract} 
\vskip 15 pt
\begin{center}
{\bf Abstract}
\end{center}
\vskip -8 pt
$\quad$
We propose a new search strategy for high-multiplicity hadronic final states.  When new particles are produced at threshold, the distribution of their decay products is approximately isotropic.  If there are many partons in the final state, it is likely that several will be clustered into the same large-radius jet.  The resulting jet exhibits substructure, even though the parent states are not boosted.  This ``accidental" substructure is a powerful discriminant against background because it is more pronounced for high-multiplicity signals than for QCD multijets.  We demonstrate how to take advantage of accidental substructure to reduce backgrounds without relying on the presence of missing energy.  As an example, we present the expected limits for several $R$-parity violating gluino decay topologies.  This approach allows for the determination of QCD backgrounds using data-driven methods, which is crucial for the feasibility of any search that targets signatures with many jets and suppressed missing energy.  
\end{abstract}

\maketitle
\newpage

\section{Introduction}
\label{Sec: Introduction}

Our approach to jet physics is undergoing a renaissance.  While most LHC studies use the energy and momentum of a jet, there is growing appreciation for the wealth of information that can be extracted by analyzing a jet's internal structure~(see \cite{Salam:2009jx,Abdesselam:2010pt,Altheimer:2012mn} for reviews).  Jet substructure gained traction when it was shown to increase the LHC sensitivity to Higgs boson decays into $b$-quarks~\cite{Butterworth:2008iy}.
Since then, jet substructure has been applied by theorists to a variety of scenarios~\cite{Butterworth:2007ke,Thaler:2008ju,Almeida:2008tp,Kaplan:2008ie,Ellis:2009su,Plehn:2009rk,Krohn:2009th,Kribs:2009yh,Plehn:2011tg,Thaler:2010tr,Gallicchio:2010sw,Gallicchio:2010dq,Cui:2010km,Almeida:2010pa,Thaler:2011gf,Hook:2011cq,Gallicchio:2011xq,Soper:2011cr,Jankowiak:2011qa,Ellis:2012sn,Jankowiak:2012na, Han:2012cu,Soper:2012pb},
and its power has been demonstrated experimentally in Tevatron \cite{Abazov:2011vh,Aaltonen:2011pg} and LHC \cite{Brooijmans:1077731,ATLAS-CONF-2011-073,ATLAS-CONF-2012-065,ATLAS:2012am,Aad:2011kq,Aad:2012meb,Chatrchyan:2012mec} searches.

In all existing studies, jet substructure has been used to search for boosted resonances with collimated decay products that are reconstructed as a single jet.  For a typical event at the LHC, parent particles are produced near threshold; the decay products are boosted for the small fraction of signal events produced with significant transverse momentum,\footnote{For example, the signal efficiency when targeting boosted gluinos is roughly $\mathcal{O}$(few \%) at the LHC~\cite{ATLAS:2012dp,Curtin:2012rm}.} or in the case where the parent particle decays to significantly lighter daughters.  In this paper, we explore a new application for jet substructure techniques that does not rely on having collimated decay products.  

We demonstrate that substructure technology is useful in the non-boosted regime for models that yield a high multiplicity of hadronic final states.  This strategy relies on the fact that when new particles with $\OO(\text{TeV})$ masses are produced at threshold, their decay products tend to be distributed isotropically in the detector. Our proposal requires an event to contain several (specifically, four or more) large-radius jets defined using the anti-$k_T$ algorithm \cite{Cacciari:2008gp} with angular size $R=1.2$.    Because these so-called ``fat" jets can cover a large fraction of the effective detector area, several decay partons from a high-multiplicity signal will often get clustered into a single fat jet.  Non-boosted final states can therefore manifest ``accidental substructure.''

Requiring multiple fat jets with non-trivial substructure greatly reduces QCD contamination.  For an event to have four fat jets, it must have at least this many well-separated hard partons.  The presence of substructure in the remaining QCD sample is most likely to occur when one or more isolated partons undergoes a hard $1\rightarrow2$ splitting.  Because this process is dominated by a soft and/or collinear singularity, the probability decreases as the energy and separation of the final states increases.  As a result, QCD events typically have suppressed substructure. 

\begin{figure}[b]
\begin{center}
\includegraphics[width=0.89\textwidth]{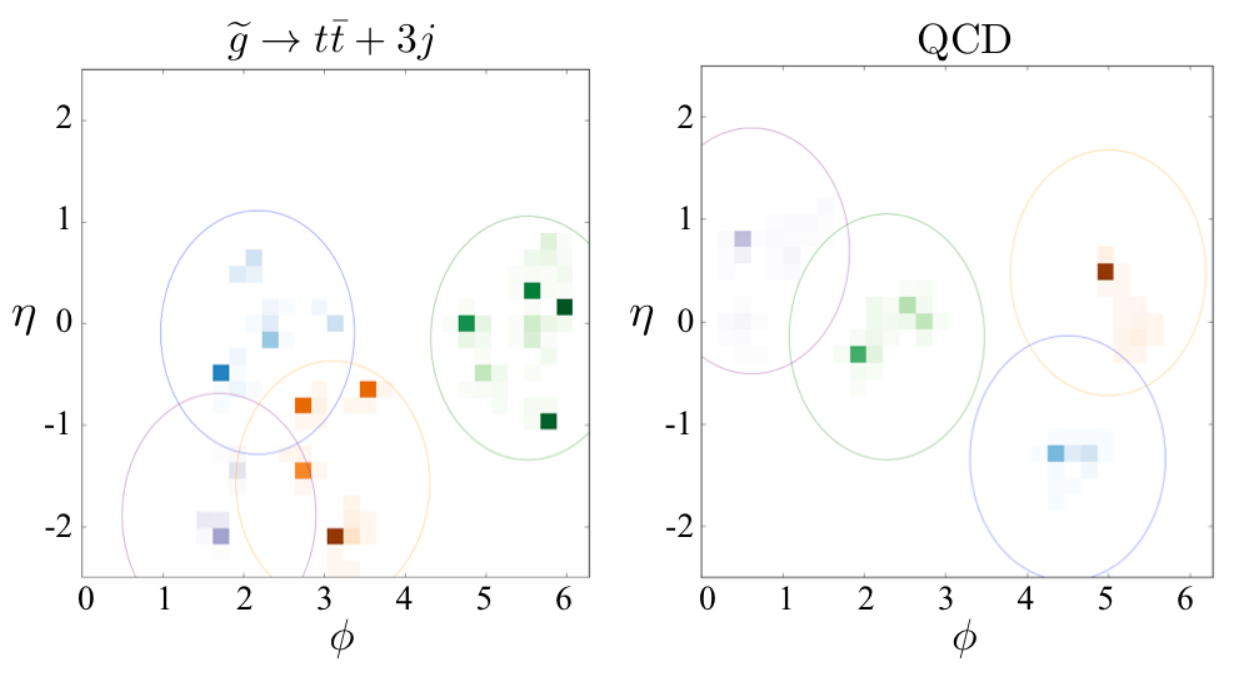}
\end{center}
\vskip -15pt
\caption{Lego plots showing the distribution of calorimeter activity in the $\eta-\phi$ plane.  The different colors correspond to different fat jets; within each panel, darker colors signify higher $p_T$ in a given detector cell.  Note that the relative $p_T$ scale is different for the signal and background example.  The signal (left panel) is pair production of 500 GeV gluinos with $\widetilde{g} \rightarrow t\,\overline{t}+3\,j$, which yields up to 18 partons in the final state.  The gluinos have transverse momenta of 120 and 65 GeV, so they are essentially at rest.  A QCD multijet event is depicted in the right panel.  The circles are centered on the clustered fat jet with a radius of $R=1.2$ to schematically illustrate the extent of each fat jet.  There is significant substructure for the signal and suppressed substructure for the background.}
\label{fig:Lego}
\end{figure}

\begin{figure}[tb]
\begin{center}
\includegraphics[scale=0.35]{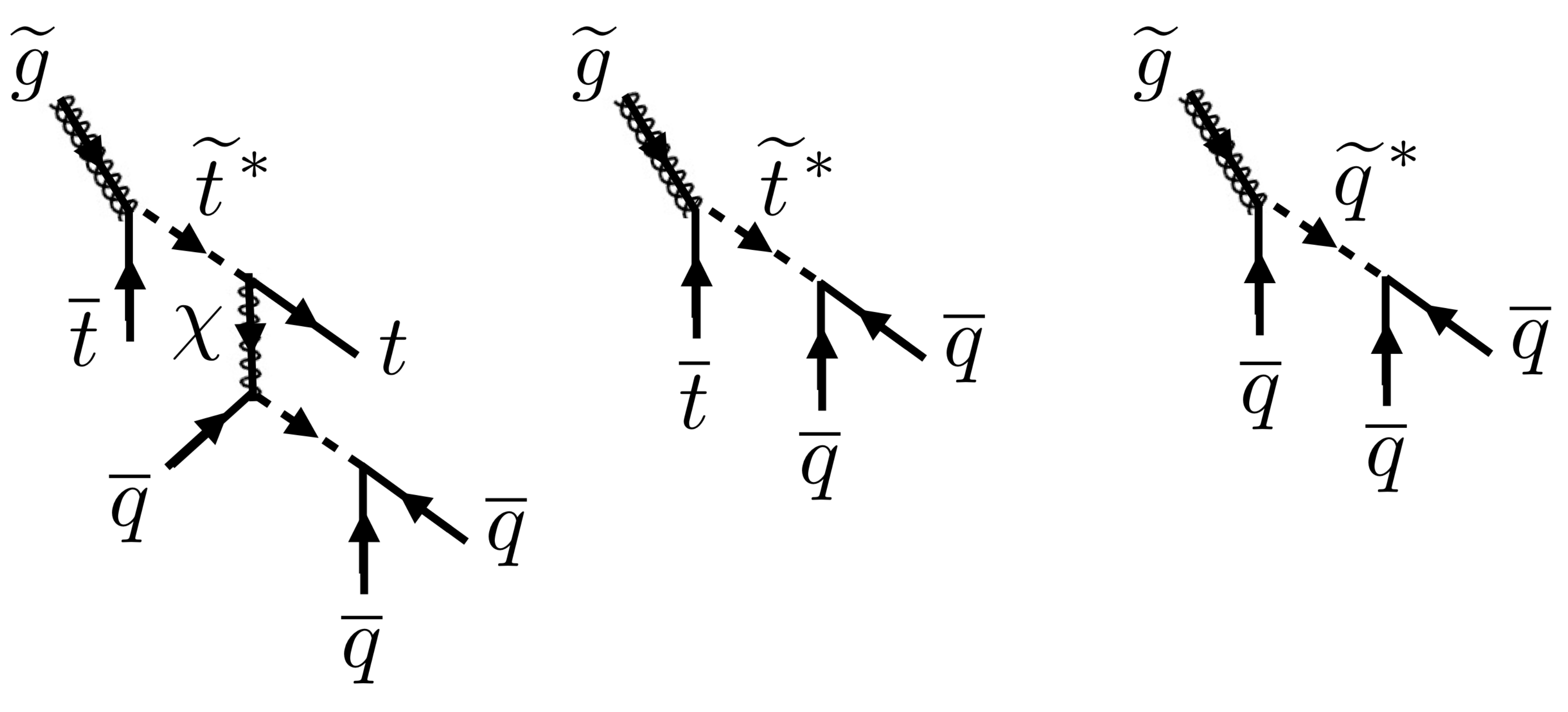}
\end{center}
\scalebox{1.45}{$\!\!\!\left[\,\widetilde{g} \rightarrow t\,\overline{t}+3\,j\,\right]\quad\quad\,\,\left[\,\widetilde{g} \rightarrow t+2\,j\,\right]\quad\quad\quad\left[\,\widetilde{g} \rightarrow 3\,j\,\right]$}
\caption{Gluino decay diagrams, illustrating topologies that can lead to as many as 18, 10, and 6-parton final states (left to right, respectively) when the gluinos are pair-produced.  Note that $\widetilde{g}$ is a gluino, $\widetilde{t}$ is a stop, $t$ is a top quark, $\widetilde{q}$ is a first or second generation squark, $\chi$ is a neutralino, and $j$ refers to a final state quark or anti-quark.}
\label{fig:decayDiagrams}
\end{figure}

Figure~\ref{fig:Lego} illustrates why accidental substructure is useful for distinguishing between a typical signal and background event.  These ``lego plots" show the spatial distribution of calorimeter activity in the $\eta-\phi$ plane, where $\eta$ is pseudorapidity and $\phi$ is azimuthal angle.  The left panel is a lego plot for a signal event with up to 18 partons in the final state; the signal is gluino pair production, where each gluino decays to a pair of top quarks and an unstable neutralino that decays to three partons (see the left diagram in Fig.~\ref{fig:decayDiagrams}).  The right panel shows the lego plot for a QCD event.  The different colors correspond to different fat jets in the event.  It is clear that the fat jets from signal have more pronounced substructure than the ones from QCD.

Figure \ref{fig:Lego} suggests that cutting on the number of small-radius ($R\sim0.4$) jets may suffice to distinguish signal from background.  An explicit high jet multiplicity search requires accurate modeling of the QCD background, which has intrinsic theoretical challenges.  The current state of the art is tree-level QCD calculations that rely on matrix element-parton shower matching up to six jets.  Because additional jets must be generated by the parton shower, these calculations systematically underestimate the $p_T$ spectrum of the high multiplicity tail.  Higher multiplicity, matched, next-to-leading order calculations are not anticipated in the near future, implying that precision modifications to the shapes of the QCD distribution will not be known.  Finally, even once this has been achieved, there is the computational limitation associated with populating the entire 3-$n$ dimensional phase space for events with $n$ jets.  As a result, theorists should validate Monte Carlo background predictions against data to derive plausible limits.  There exist studies from the CMS and ATLAS collaborations that present 6 jet \cite{:2012gw, ATLAS:2012dp} and 8 jet \cite{CMS-PAS-EXO-11-075} distributions.  However, these do not provide enough information to place cuts on the number of small-radius jets larger than $\sim$ 6--8.  This constrains theoretical investigations of high multiplicity searches with small-radius jets.  

An experimental analysis targeting many small-radius jets must obtain the multijet backgrounds from data.  Current data-driven methods for determining detailed kinematic features of small-radius jets are limited in that they rely on ad hoc fitting functions to perform background extrapolations.  If a search that utilized these procedures yields an excess of events, there is no guidance for investigating the discrepancy because the functions are not derived from an underlying theory.\footnote{For recent theoretical progress on extrapolating jet multiplicity, see \cite{Gerwick:2012hq, Gerwick:2012fw}.}  

Searches that use fat jets can implement an alternate strategy to estimate backgrounds.  For the substructure analysis proposed here, one can study the internal structure of fat dijets.  Because this sample should be signal poor, it can be used to determine the pure QCD dependence of jet mass and substructure on other quantities like jet $p_T$.  These results can then be extrapolated to four fat jet events, and should lead to reasonable background predictions so long as the correlations between fat jets are small.  Importantly, the associated systematics for a fat jet analysis differ from those that dominate in a search for many small-radius jets.  It is beneficial to have competing searches with different systematics to ensure that new physics is not overwhelmed by large uncertainties.

Finally, we note that our analysis does not rely on the presence of missing transverse energy (\MET$\!\!$), which is typically crucial for discriminating against multijet backgrounds in searches for supersymmetry (SUSY).  Missing energy is not a robust prediction of SUSY models, \emph{e.g.}  $R$-parity can be violated, the superpartner spectrum can be squeezed, or SUSY can be stealthy \cite{Fan:2011yu, Fan:2012jf}.  There are also a number of non-SUSY models that have signatures without \MET$\!\!$, such as \cite{Kilic:2009mi,Gross:2012bz, Kang:2008ea, Cheung:2008ke}.  To cover these and other \MET$\!\!$-less theories at the LHC, it is imperative to develop new search strategies to efficiently reduce the QCD background.  Such a strategy could rely on rare objects in the event, such as $b$-jets or leptons, to further reduce backgrounds.  However, a search that is independent of these extra handles is powerful for its generality.  Because our proposal \emph{only} relies on having a final state with many jets, it can be used to place limits on a wide-range of model space.  

We demonstrate that accidental substructure is a powerful discriminator by applying it to three distinct gluino $\go$  decay scenarios when the $R$-parity violating (RPV) superpotential coupling $U^c\,D^c\,D^c$ is non-zero:
\begin{equation}
 \go\rightarrow t \,\bar{t} + 3\,j,  \quad \quad \go\rightarrow t + 2\,j,  \quad \text{ and } \quad \go\rightarrow 3\,j.
\end{equation}
Here $j$ refers to a final state quark or anti-quark, not to a detector-level jet.  When the gluinos are pair-produced, these three topologies can lead to as many as 18, 10, and 6-parton final states, respectively, as shown in Fig.~\ref{fig:decayDiagrams}.  The first topology arises when a gluino decays to a pair of tops and an unstable neutralino, which decays to three partons through an off-shell squark via $U^c\,D^c\,D^c$.  The other two topologies correspond to the RPV gluino decays into $t\,b\,s$ and $u\,d\,s$ final states.   For a review of constraints on these RPV interactions, see~\cite{Barbier:2004ez}.  The 18 and 10-parton topologies are particularly well-motivated theoretically because the top quarks in the final state can result from a light stop in the spectrum.  This is a plausible scenario with minimal fine-tuning where the non-zero RPV couplings suppress \MET$\!\!$, thereby hiding SUSY from current searches~\cite{Brust:2012uf}.  In particular, the 10-parton topology was the focus of a recent proposal that used substructure techniques to look for boosted stops \cite{Han:2012cu}.

The remainder of this paper proceeds as follows. In Sec.~\ref{section: analysisstrategy}, we present the needed variables, jet mass and $N$-subjettiness, and introduce the concept of ``event-subjettiness."  In Sec.~\ref{section:examples}, we show how these tools can be combined into a full analysis.  After a brief description of the event generation procedure, we present the expected limits for the different gluino decay topologies.  We conclude in Sec.~\ref{section:conclusion}.  Appendix~\ref{sec:Validation} contains a detailed description of our simulations, including validation plots.     

\section{Quantifying Accidental Substructure}
\label{section: analysisstrategy}

Our analysis relies on two observables: total jet mass and event-subjettiness.  The latter is a new variable that we introduce to quantify the amount of accidental substructure in an event.  It requires $N$-subjettiness to characterize the subjet nature of each jet.  Jet mass, $N$-subjettiness, and event-subjettiness form the cornerstone of our analysis, so we introduce them individually here.  The full analysis strategy is presented in Sec.~\ref{section:examples} and the details of our Monte Carlo event generation, detector mock-up, and validation can be found in Sec.~\ref{section: MonteCarlo} and Appendix~\ref{sec:Validation}.  

For the figures in this section, we select 8 TeV LHC events with at least four jets, clustered using the anti-$k_{T}$ algorithm~\cite{Cacciari:2008gp} with cone size $R=1.2$.  The transverse momenta of the leading and subleading fat jets must satisfy $p_{T}\ge100$ GeV and $p_{T}\ge 50$ GeV, respectively.  Although no 8 TeV multijet, \MET$\!\!$-less triggers are publicly available, some 7 TeV examples include:   five or more jets ($R=0.4$) with $p_T > 30$ GeV at ATLAS \cite{ATLAS:2012dp}, $\sim500$--$750$ GeV of $H_T$ at CMS \cite{CMS-PAS-EXO-11-075}, and 4, 6, or 8 high-$p_T$ jets ($R=0.5$) at CMS~\cite{CMS-PAS-SUS-12-009}.  We have verified that the first of these triggers is 100\% efficient for the QCD background and the gluino topologies we consider after final selection cuts.

\subsection{Jet Mass}
\label{section:jetmass}

Standard SUSY searches at ATLAS and CMS use a combination of missing energy, \MET$\!\!$, and visible transverse energy, 
\be
H_T = \sum_{j = 1}^{N_j} \sqrt{(p_T^2)_j+m_j^2},
\ee  
where $j$ is a jet in the event with mass $m_j \equiv \sqrt{E_j^2 - |\vec{p}_j|^2}$ and $N_j$ is the number of jets in the event with $p_T > 50 \GeV$.  The total jet mass of an event,  
\be
M_J \equiv \sum_{j = 1}^{N_j} m_j,
\ee
is a more powerful discriminator than $H_T$ in searches for high multiplicity final states~\cite{Hook:2012fd} because a jet's mass automatically encodes gross kinematic features of its constituents.  

\begin{figure}[t!]
\centering
\includegraphics[scale=0.35]{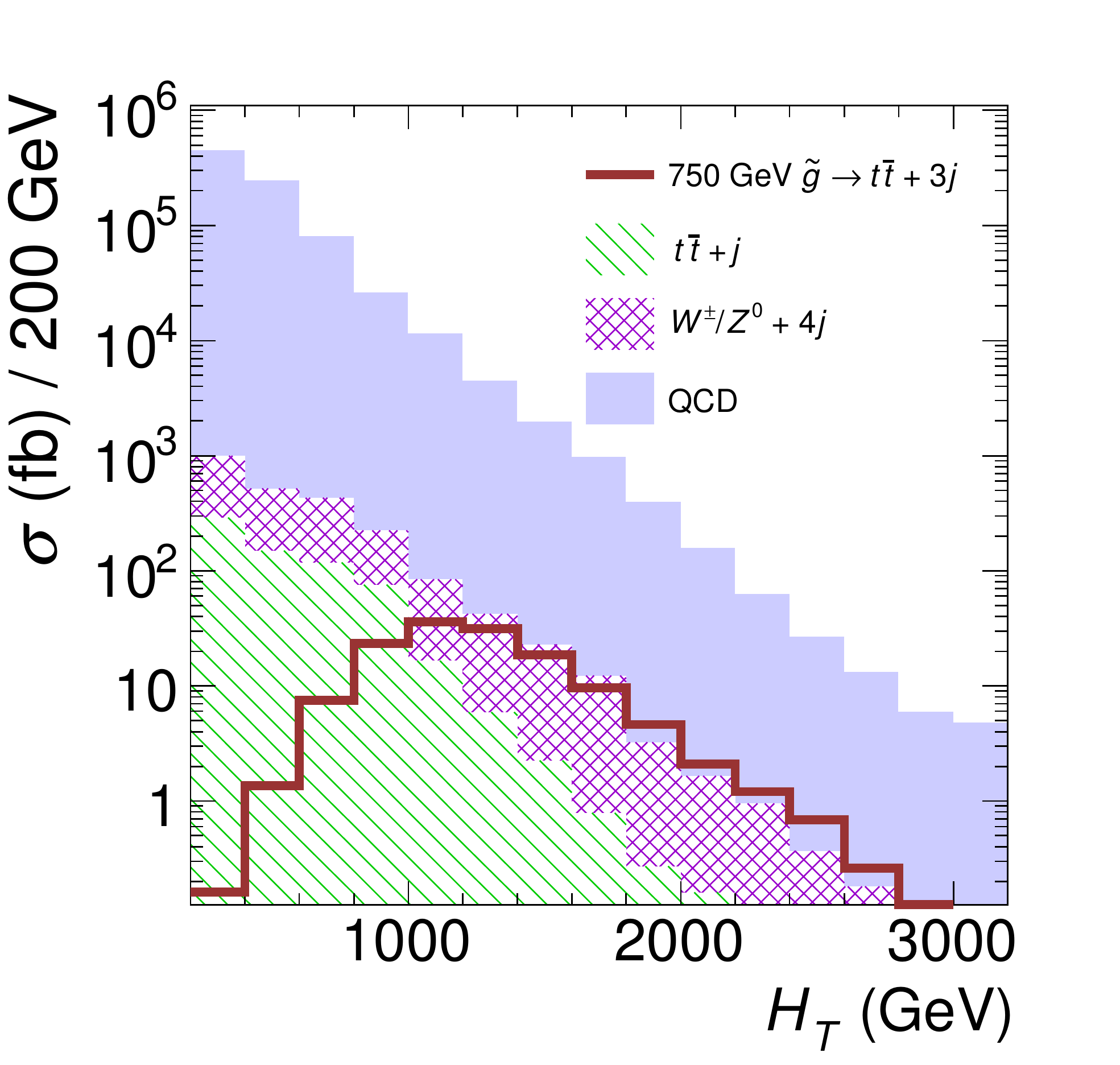}
\includegraphics[scale=0.35]{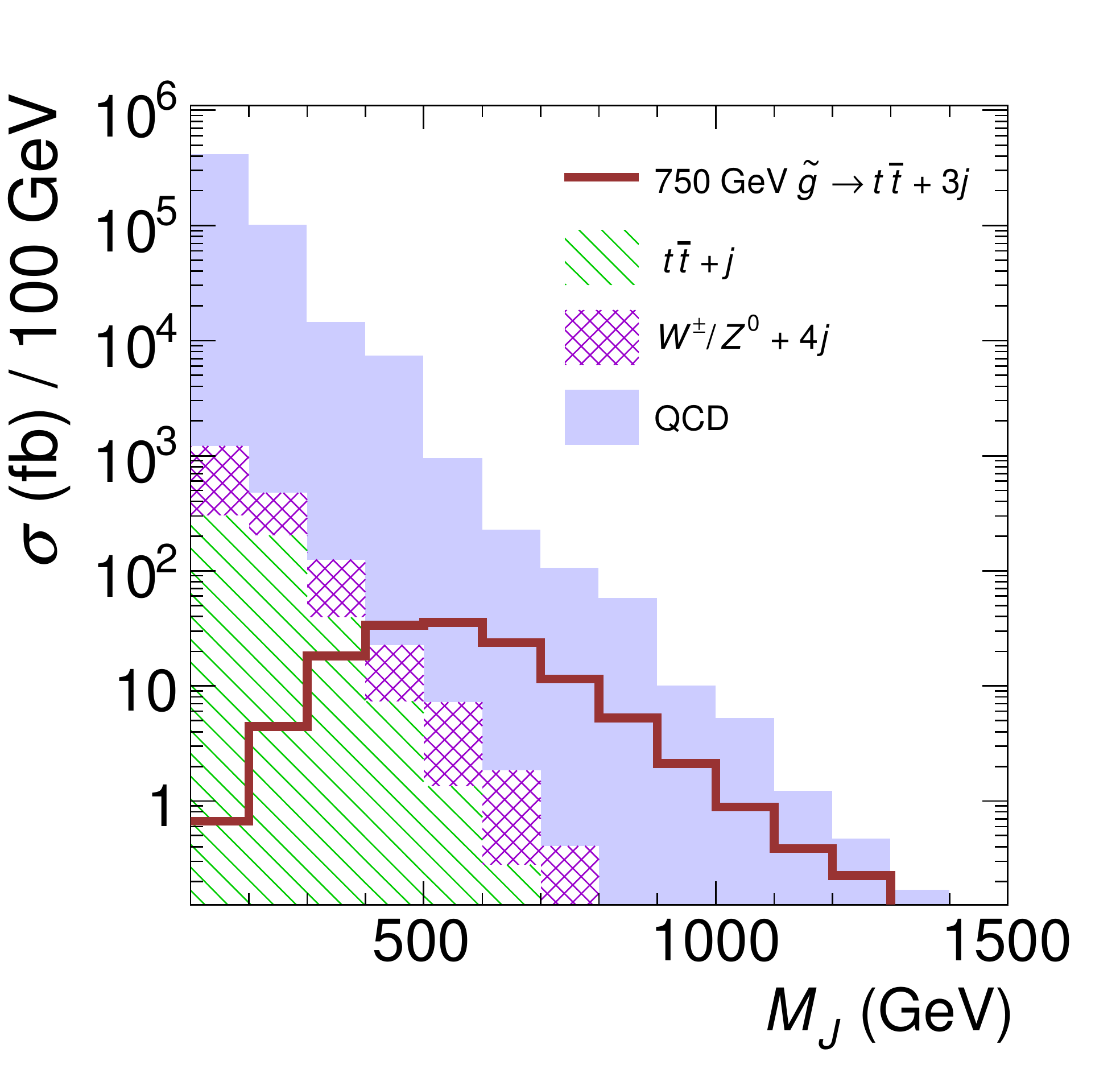}
\caption{The $H_T$ (left) and $M_J$ (right) distributions for the backgrounds and an example signal.  The signal (red solid line) is pair production of a $750 \mbox{ GeV}$ gluino with $\widetilde{g} \rightarrow t\,\bar{t}+3\,j$.  The stacked histogram is for background (QCD in solid blue, $W^\pm/Z^0 + 4\,j$ in hatched magenta, and $t\,\bar{t}+j$ in striped green).  $M_J$  is a more powerful discriminator than $H_T$ when comparing signal to background.}
\label{fig:compareHTandMJ}
\end{figure}

Consider a small-radius jet that is seeded from an isolated parton.  In the absence of showering, this jet will have zero mass.  Non-zero jet mass arises if multiple partons are clustered together and/or from QCD radiation --- the former yields a larger jet mass than the latter.  As a result, a QCD and signal event with equivalent $H_T$ can have different total jet mass.  More quantitatively, $H_T$ can be related to $M_J$ via
\begin{eqnarray}
H_{T} &=& \sum_{j = 1}^{N_j} \sqrt{(p_{T}^2)_j + m_j^2}
\mbox{  } \propto \mbox{  } \sum_{j=1}^{N_j} \sqrt{\langle m_j^2 \rangle((\kappa R)^{-2} +1)} \mbox{  } \simeq \mbox{  } M_J \frac{\sqrt{ 1+ (\kappa R)^2}}{ \kappa R},
\label{eq: HT}
\end{eqnarray}
where $\kappa \simeq \sqrt{\alpha_s}$ for jets whose mass is generated from the parton shower~\cite{Salam:2009jx} and $\kappa \simeq 1$ for fat jets that contain multiple hard partons accidentally clustered in the same jet.  Figure~\ref{fig:compareHTandMJ} shows the $H_T$ and $M_J$ distributions for background and a signal example.  Clearly, a cut on $M_J$ improves sensitivity to the signal as opposed to an $H_T$ requirement.

The authors of~\cite{Hook:2012fd} proposed a study that took advantage of total jet mass for high multiplicity signals, but which still relied on a missing energy cut.  In this work, we demonstrate that accidental substructure increases sensitivity when used in conjunction with total jet mass. This result is especially useful in topologies with \MET suppression, such as the benchmarks presented in Fig.~\ref{fig:decayDiagrams}. Adding a moderate \MET cut for other topologies that do contain sources of missing energy, {\it e.g.} new physics signals with tops in the final state, can provide an additional handle for improving the discriminating power of accidental substructure and jet mass.

\subsection{$N$-subjettiness}
\label{subsection:nsubjettiness}

To quantify accidental substructure, we begin by considering the $N$-subjettiness variable $\tau_{N}$~\cite{Thaler:2010tr,Thaler:2011gf,Stewart:2010tn}. $\tau_N$ is a measure of the degree to which a fat jet has $N$ well-separated subjets.  For each jet, $\tau_N$ is defined as
\begin{align}
\tau_{N}&=\frac{1}{d_{\beta}}\sum_{i}(p_T)_i\min\left\{
\Delta R^\beta_{i,1}, \Delta R^\beta_{i,2},..., \Delta R^\beta_{i,N}
\right\}\notag\\
d_\beta&=\sum_{i}(p_{T})_i \, R^\beta_0,
\end{align}
where the minimization is performed by varying $N$ axes, $R_0$ is the choice of clustering radius, and $\Delta R_{i,M} = \sqrt{(\Delta \phi_{i,M})^2 + (\Delta \eta_{i,M})^2}$ denotes the angular distance between the $i^\textrm{th}$ constituent particle and the $M^\textrm{th}$ axis. We take $\beta = 1$ and $R_0=1.2$.

To elucidate what $N$-subjettiness measures, consider $\tau_3$.  If the jet consists of three or fewer well-collimated subjets, $\tau_3 \simeq 0$ because $\mbox{min}\{\Delta R_{i,1},\Delta R_{i,2}, \Delta R_{i,3}\}$ vanishes for the $i^{\text{th}}$ constituent.  If the fat jet contains more than three subjets (or the particles making up the jet are not well-collimated), $\tau_3 > 0$ because at least one subjet is not aligned with an axis.   

While the individual $\tau_N$ are not typically useful, ratios are~\cite{Thaler:2010tr}.  For example, 
\be
\tau_{NM} \equiv \tau_N/\tau_M
\ee
is efficient at selecting $N$-subjetty events for $M < N$.  For a jet with $N$ well-separated subjets, $\tau_M$ is large, $\tau_N$ is small, and therefore $\tau_{NM}$ is much less than 1.  Rejecting events with $\tau_{NM} \sim 1$ selects for jets that are more $N$-prong like.\footnote{Ensuring that the $\tau_{NM}$ variables are infrared and collinear (IRC) safe~\cite{Soyez:2012hv} is important for the implementation of our proposed search.  $\tau_{N(N-1)}$ is IRC safe if applied to a jet with $N$ hard prongs, which is ensured by requiring a lower bound on $\tau_{(N-1)(N-2)}$.  While the results presented here do not include this requirement, we have verified that they are not significantly changed by a naive application of the IRC safety conditions.}

\begin{figure}[b]
\centering
\includegraphics[scale=0.7]{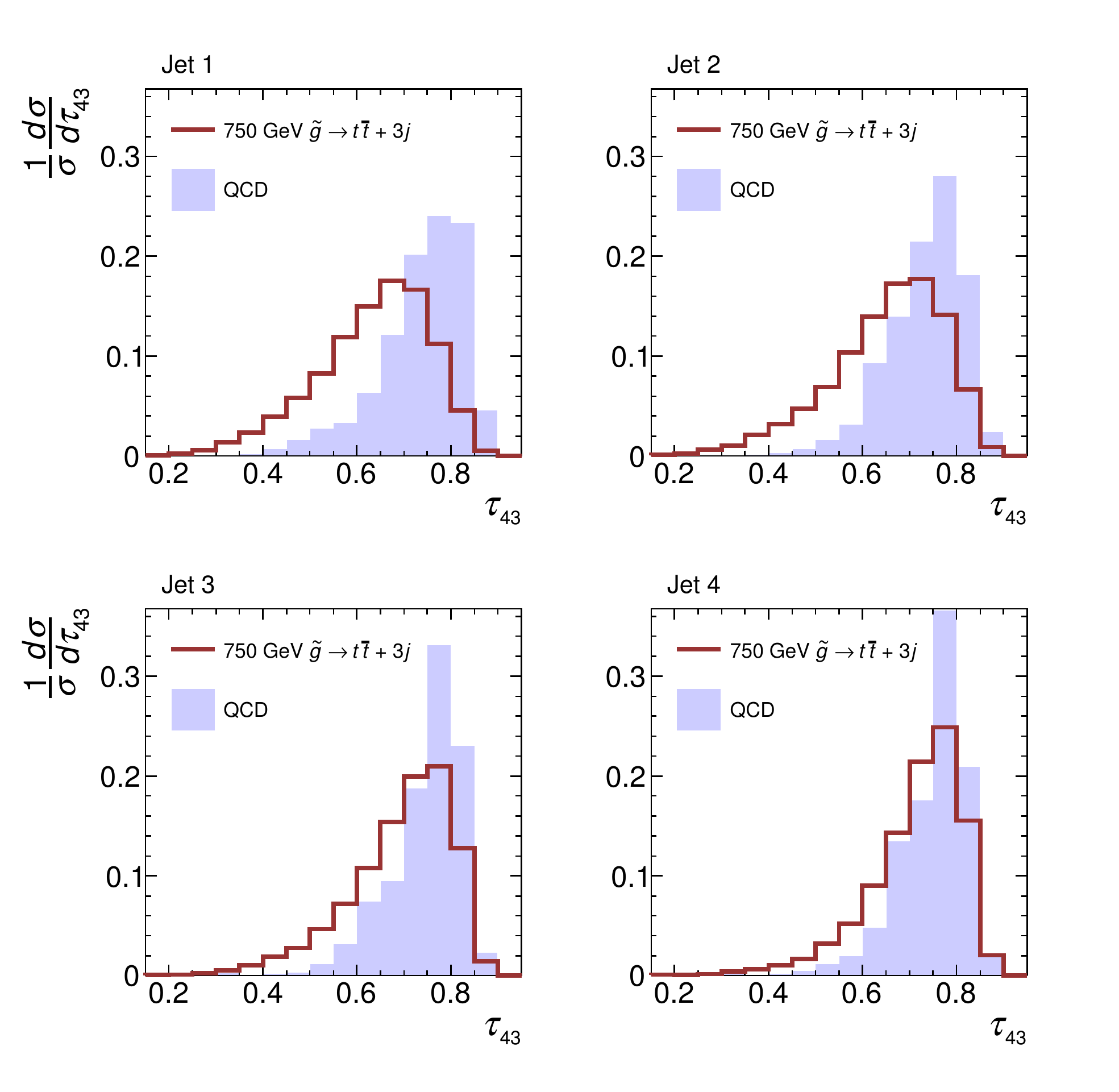}
\caption{Normalized distributions of $\tau_{43}$ for background and a signal example.  Each plot shows the normalized distribution before a cut on $M_J$.  The signal (red solid line) is pair production of a $750 \GeV$ gluino with $\widetilde{g} \rightarrow t\,\bar{t}+3\,j$.  The solid blue histogram is for the QCD background.    Each panel is the distribution for the $j^\mathrm{th}$ jet; the order is by decreasing $p_T$.  Note that the top and electroweak backgrounds are subdominant and are not shown here.}
\label{fig:tau43}
\end{figure}

Figure~\ref{fig:tau43} shows the normalized distributions of $\tau_{43}$ for each of the four hardest jets for QCD and the $\widetilde{g} \rightarrow t\,\bar{t}+3\,j$ topology.  The jets in each event are ordered by decreasing $p_T$.  The background sample is peaked around $\tau_{43} \sim 0.7$--$0.8$.  In contrast, the distribution for the signal is shifted to lower values, with a tail that extends to lower $\tau_{43}$.  These distributions reinforce the general conclusions we drew from the lego plots in Fig.~\ref{fig:Lego}.  Specifically, $\tau_{43}$ is shifted towards lower values for the signal relative to the background, suggesting that signal jets typically look more four-subjetty than the background jets.    

\subsection{Introducing Event-subjettiness}

$N$-subjettiness is useful for characterizing the number of subjets in a single jet.  However, it would be useful to have a variable that takes into account the relative abundance of jets with substructure in an entire event.  To this end, we introduce ``event-subjettiness," $T_{NM}$, which is defined as the geometric mean of the $\tau_{NM}$ for the four hardest jets in an event:  
\begin{equation}
T_{NM} =  \Bigg[ \prod_{j = 1}^4 \Big(\tau_{NM}\Big)_{j} \Bigg]^{1/4}.
\end{equation}
The more jets with substructure in an event, the more jets with a small $\tau_{NM}$, resulting in a smaller value of $T_{NM}$.  The geometric mean is less sensitive to the presence of a single high $\tau_{NM}$ in an event than the arithmetic mean.  In particular, the arithmetic (geometric) mean tends to result in slightly larger $S/B$ ($S/\sqrt{B}$) than the geometric (arithmetic) mean.  This leads to a mild improvement in the reach when using the geometric mean.  We also explored placing cuts on combinations of the $\tau_{NM}$ for the single two hardest jets; this does not lead to the same level of discriminating power because the amount of substructure is not necessarily correlated with the hardness of a jet.  As discussed in Sec.~\ref{section:jetmass}, two jets with equivalent $p_T$ can have a different mass depending on whether the mass arises from accidental substructure or hard-emission.

\begin{figure}[t]
\centering
\includegraphics[scale=0.35]{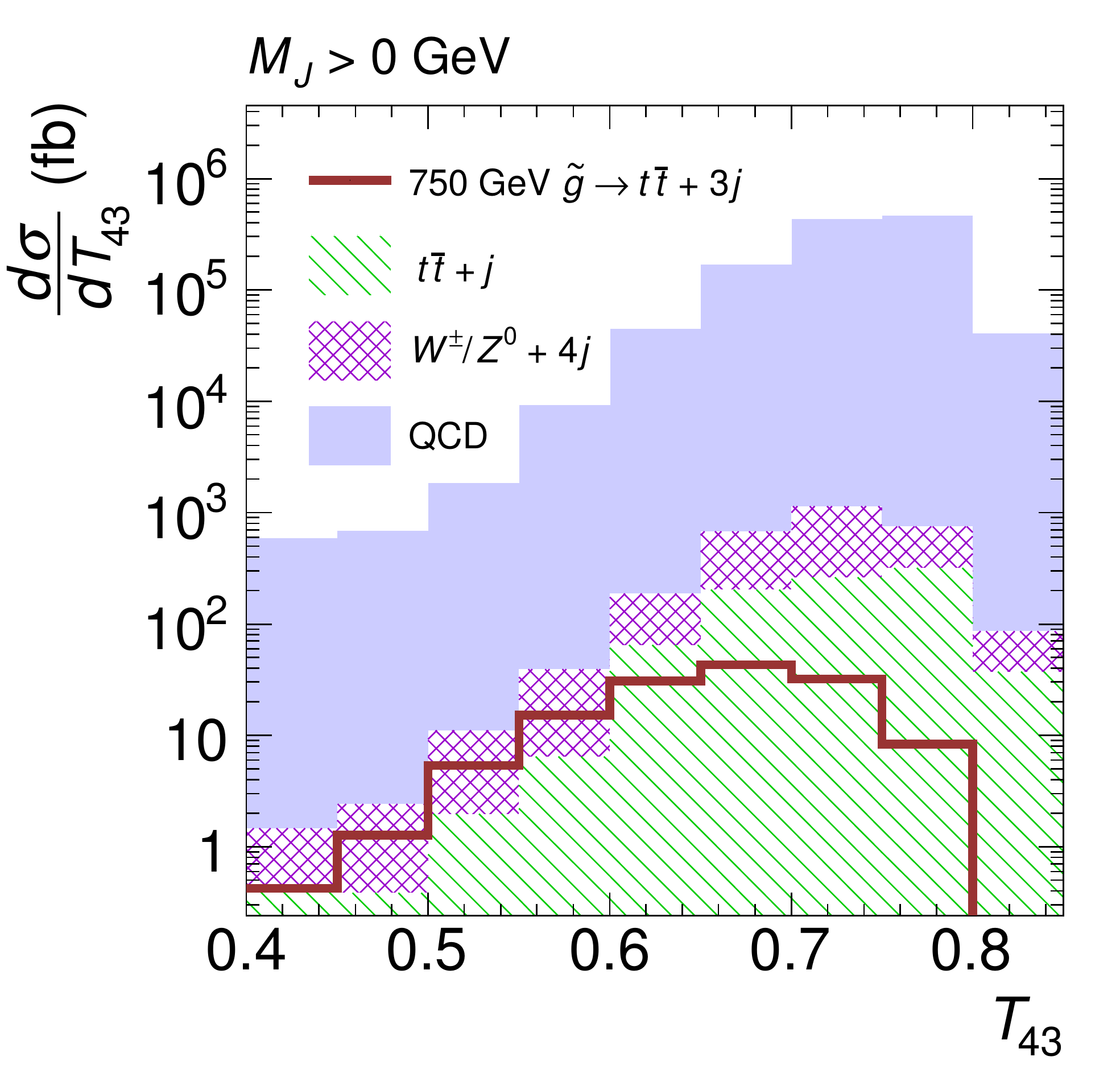}\includegraphics[scale=0.35]{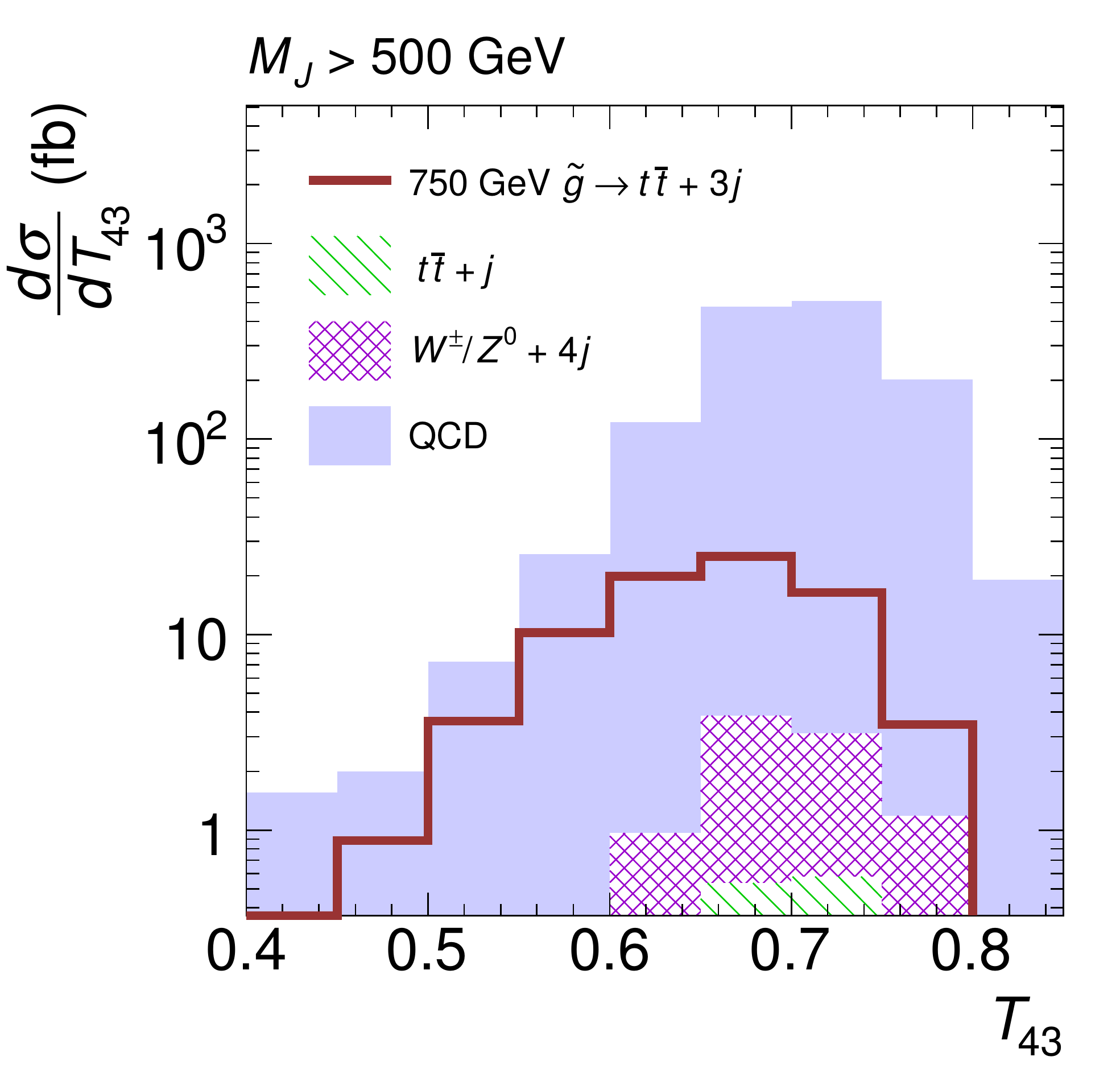}
\caption{Distributions of $T_{43}$ for backgrounds and an example signal, with $M_J>0$ (left) and $M_J>500 \GeV$ (right).  The signal (red solid line) is pair production of a $750 \GeV$ gluino with $\widetilde{g} \rightarrow t\,\bar{t}+3\,j$.  The stacked histogram is for background (QCD in solid blue, $W^\pm/Z^0 + 4\,j$ in hatched magenta, and $t\,\bar{t}+j$ in striped green).  A cut on $T_{43} \lesssim 0.6$ helps to distinguish signal from background, after requiring $M_J > 500 \GeV$.}
\label{fig:T43}
\end{figure}

Figure~\ref{fig:T43} illustrates the distributions of $T_{43}$ for backgrounds and the signal example with $\widetilde{g} \rightarrow t\,\bar{t}+3j$.  For this topology, many of the signal fat jets often have four or more subjets, which drives down $T_{43}$ relative to that for the backgrounds.  This is evident, for example, in Fig.~\ref{fig:Lego} where the signal event has $T_{43} = 0.45$ compared to $0.73$ for the QCD event.  As Fig.~\ref{fig:T43} shows, after a cut on the total jet mass (right panel), the ratio of signal to background improves relative to no total jet mass cut (left panel).  The right panel suggests that the signal and background can be distinguished by applying an additional cut $T_{43} \lesssim 0.6$.  We  demonstrate the efficacy of this strategy in the following section when we estimate the sensitivity to the signal topologies in Fig.~\ref{fig:decayDiagrams}.  

\section{Analysis Strategy}
\label{section:examples}

Having presented the individual components of our analysis, we now combine them and present the complete search strategy.  To illustrate the effectiveness of this approach, we compute expected limits for the three different RPV gluino decay chains in Fig.~\ref{fig:decayDiagrams}.  Of course, our proposal is quite general and can be applied to any high-multiplicity final-state.   

\subsection{Event Generation}
\label{section: MonteCarlo}

We begin by briefly describing the generation of signal and background events.  Appendix~\ref{sec:Validation} contains a more detailed description of the detector mockup and Monte Carlo validation.

QCD is the dominant background for a multijet signal with no missing energy.  \texttt{Sherpa}  1.4.0~\cite{Gleisberg:2008ta, Krauss:2001iv, Schumann:2007mg, Gleisberg:2008fv, Hoeche:2009rj} is used to generate and shower $\sim 400$ million inclusive $p\,p\rightarrow n\,j$ events, where $n\in (2,\dots,6)$.  Matrix elements for up to 6 partons are generated, which are then matched to the parton shower using the CKKW procedure~\cite{Catani:2001cc}.  All Sherpa events are generated using the default CTEQ 6.6 parton distribution function~\cite{Nadolsky:2008zw} and include the effects of underlying event.  We generated a sample of weighted events in order to increase the statistical power of our finite sample.  The Monte Carlo error, $\epsilon_{\text{MC}}$, after cuts is 
\begin{equation}
\epsilon_{\text{MC}} = \frac{\sqrt{\sum_i w_i^2}}{\sum_i w_i},
\end{equation} 
where $w_i$ is the weight of the $i^{\text{th}}$ event in the sample.  We verify that the Monte Carlo error is less than the systematic error for the signal regions of interest.

For consistency, \texttt{Sherpa} is also used to generate additional subleading background contributions.  In particular, we generate $\sim 25$ million matched and weighted $t\,\bar{t}+n_t\, j$ events, where the tops are forced to decay hadronically.  We also simulate $\sim 25$ million matched and weighted data sets for each electroweak background: $W^+ + n_W\, j$, $W^- + n_W\, j$, and $Z^0 +n_Z\, j$, where the gauge bosons are forced to decay to quarks.  Here, $n_t\in{0,1}$ and $n_W,\,n_Z\in{1,2,3,4}$.  Table~\ref{tab: cuts} shows that these non-QCD backgrounds are subdominant.  This would not be the case if a \MET cut were also applied. 

The matrix elements for gluino pair production are generated in \texttt{MadGraph5} 1.4.8.4~\cite{Alwall:2011uj} for the $\go\rightarrow t\,\bar{t}+3\,j$ topology.  Those for the $\go\rightarrow t\,+2\,j$ and $\go\rightarrow 3\,j$ topologies are generated directly in \texttt{Pythia} 8.170~\cite{Sjostrand:2000wi,Sjostrand:2002ip,Desai:2011su}, where the RPV gluino is allowed to hadronize before decaying.  All three signal topologies are generated using the default CTEQ6L1 PDF set~\cite{Pumplin:2002vw,Stump:2003yu} and are showered and hadronized in \texttt{Pythia} including the effects of underlying event.  Because the gluinos are produced at threshold and decay to several fairly hard jets, it is not necessary to perform matching. 

Both signal and background events are passed through our own detector mockup, which only includes  the effects of detector granularity.  \texttt{FastJet} 3.0~\cite{Cacciari:2005hq,Cacciari:2011ma} is used to cluster events into anti-$k_T$~\cite{Cacciari:2008gp} jets with $R=1.2$.  Variables such as jet mass and substructure are sensitive to soft, diffuse radiation that results from underlying event and pile-up.  The ATLAS study in~\cite{ATLAS:2012am} explicitly demonstrated that the mean jet mass for anti-$k_T$ jets with $R=1.0$ and $p_T>300 \GeV$ is constant with respect to the number of pile-up vertices for 35 pb$^{-1}$ of 7 TeV data, after a splitting/filtering procedure is applied. For variable multiplicity fat jets, which is quite typical for accidental substructure, filtering is not the optimal grooming technique because it places a fixed requirement on the number of subjets within the fat jet \cite{Butterworth:2008iy}.  Instead, to reduce the contamination due to soft radiation resulting from underlying event, we apply the trimming procedure of \cite{Krohn:2009th} to the jets before applying any kinematic cuts.  We require any subjets of radius $R=0.3$ to have a $p_T$ greater than 5\% of the fat jet's transverse momentum.  This choice of parameters is motivated by a recent ATLAS analysis~\cite{ATLAS-CONF-2012-065}.  We find that trimming eliminates the dependence on the different underlying event models used by the generators. 

\texttt{Prospino} 2.1~\cite{Beenakker:1996ch} is  used to obtain the NLO production cross section for the gluinos.  For the QCD background, we use a $K$-factor of 1.8, obtained by comparing distributions of the generated QCD Monte Carlo with published distributions in~\cite{Aad:2011tqa,ATLAS:2012am} (see Appendix~\ref{sec:Validation} for details on validation).  All other backgrounds are subdominant and our analysis is therefore insensitive to the exact choice of their cross sections.  We use the \texttt{Sherpa} leading order predictions for these backgrounds.

\begin{table}
\centering
\renewcommand{\arraystretch}{1}
\setlength{\tabcolsep}{7pt}
\begin{tabular}{l || c c c | c c c}
\hline
Requirement                                          	&     QCD					& $t\,\bar{t}+\,j$    	& $W^\pm/ Z^0+\,4\,j$	&  $\go\rightarrow t\,\bar{t}+3\,j$	&  $\go\rightarrow t +2\,j$ 	& $\go\rightarrow 3\,j$\\[1ex] 
\hline
(1) \,$N_j$ = 4								&	$5.8\times10^6$	&	4500					&	$1.0\times10^4$ 		&	$680$					&  $7200$						&	$4800$	\\ 
(2) \,$M_J > 500 \GeV$		\quad			&	$6800$ 				&	8.4					&	40								&	$400$					&  $990$						&	$640$	\\ 
(3) \,$T_{43} < 0.6$ \quad\,	\emph{or}	 \,	&	$180$					&	0.61					&	1.5							&	$\textbf{75}$				& $\textbf{110}$				&  	(48)	\\ 
\hspace{0.24in}$T_{21} < 0.2$  			&	$77$						&	0.047 				&	1.1							&	$(1.7)$					& $(27)$						&	\textbf{39}	\\ 

\end{tabular}
\caption{Event yields from our Monte Carlo simulation, assuming 5 fb$^{-1}$ of 8 TeV data and taking the gluino mass to be 750 GeV for $\go\rightarrow t\,\bar{t}+3\,j$ and 500 GeV for the other two topologies.  The table shows the number of events after requiring (1) four fat jets with $m_j > 20 \GeV$ and the appropriate $p_T$ requirements (see text), then (2) a cut on the total jet mass, and then (3) a cut on  event-subjettiness for a given choice of $T_{NM}$.  Yields are shown for two different $T_{NM}$ cuts that are optimized for the 18, 10, and 6-parton topologies; the number of events that corresponds to the best choice for this cut is bolded while the non-optimal choices are in parentheses.  
}
\label{tab: cuts}
\end{table}
\subsection{Expected Reach}
Now, we are ready to compute the expected reach of our analysis.  All events are required to satisfy the following criteria.  Each event must have at least four fat jets, where the $p_T$ of the hardest jet is at least 100 GeV and the $p_T$ of the next three hardest jets is at least 50 GeV.  To reduce contamination of heavy flavor resonances and high-$p_T$ QCD jets with no hard splittings, only jets with $m_j > 20 \GeV$ are considered. To further reduce QCD and $t\,\bar{t}$ background contributions, each event must have at least 500 GeV of total jet mass, $M_J$.  Finally, a cut is placed on event-subjettiness, $T_{NM}$.  The cuts for $M_J$ and $T_{NM}$ were selected to maximize significance, while ensuring that the Monte Carlo error remained below the systematic error.  This requirement imposes a significant limitation on our ability to fully optimize the search and is the reason we only present one set of cuts.  Table~\ref{tab: cuts} summarizes the cut efficiencies on signal and background. 

To determine the expected reach for the three topologies in Fig.~\ref{fig:decayDiagrams}, we assume that the probability of measuring $n$ events is given by the Poisson distribution with mean $\mu = B+S$, where $B$ and $S$ are the number of expected background and signal events, respectively.  The probability of measuring up to $N_m$ events is
\begin{equation}
P\left(N_m |  \mu \right) = e^{-\mu} \sum_{n=0}^{N_m} \frac{\mu^n}{n!}. 
\label{eq: poisson}
\end{equation}
This expression assumes that there is no uncertainty in the value for $B$.  In the presence of a systematic uncertainty $\sys$, \Eref{eq: poisson} must be convoluted with the probability distribution of $B$, which we assume is log-normal because $B\ge0$:
\begin{equation}
P_{\text{sys}}\left(N_m, S, B\right)  = \int_0^{\infty} \text{d}x \text{ }  P\left(N_m | S + x\right) \cdot \ln \mathcal{N}\left( x \right),
\label{eq: conv}
\end{equation}
where $\ln\mathcal{N}(x) = \frac{1}{x \sqrt{2 \pi} \sys}\text{exp}\Big[- \frac{( \ln x - \ln B )^2}{2\epsilon_{\text{sys}}^2}\Big]$.  Note that as $\sys \rightarrow 0$, the log-normal distribution becomes a delta function centered at $B$ and \Eref{eq: conv} reproduces the standard result for a Poisson distribution.  To obtain the expected limit on the signal cross section, we solve \Eref{eq: conv} for $S$ assuming that $N_m = B$ and $P_{\text{sys}} = 0.05$ (95\% exclusion).  We find that the expected limits are not sensitive to the distribution function chosen for $B$; a Gaussian distribution gives essentially the same result.     

An ATLAS analysis of the full 2011 dataset reported a jet mass scale systematic uncertainty of $\sim 4\text{--}8\%$ (depending on jet $p_T$) for anti-$k_T$ trimmed jets with $R=1.0$~\cite{ATLAS-CONF-2012-065}.  For four fat jets, this gives at most a 16\% systematic uncertainty.  To be conservative and to account for additional sources of systematic effects (\eg ~jet energy scale), we take $\sys = 20\%$ when computing sensitivities.  
\begin{figure}[tb]
\centering
\includegraphics[scale=0.4]{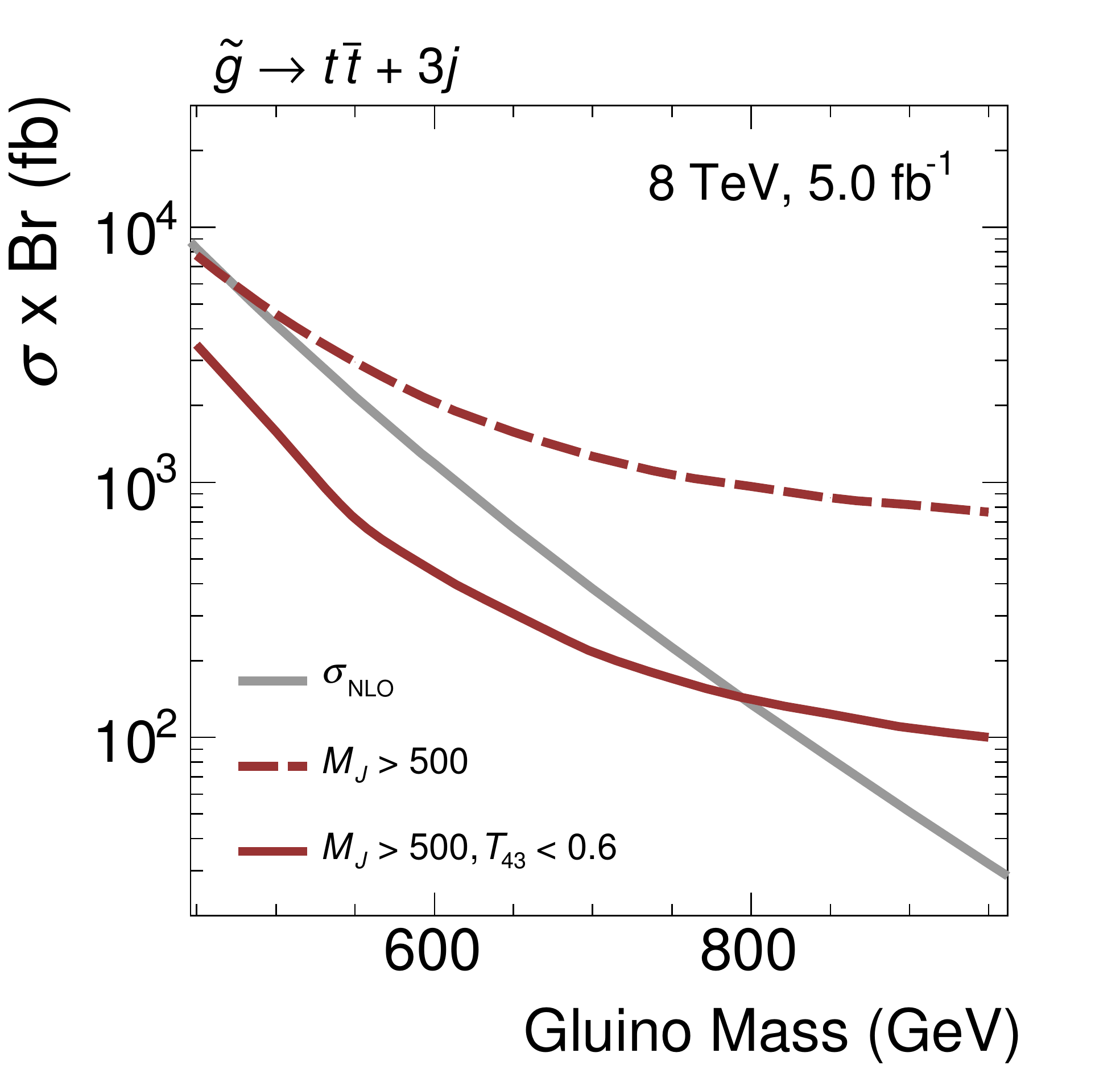}
\caption{The 95\% expected exclusion curves for the $\widetilde{g} \rightarrow t\,\overline{t}+3\,j$ topology at the 8 TeV LHC with 5 fb$^{-1}$ of data.  The solid grey curve is the NLO prediction for the gluino pair production cross section computed using \texttt{Prospino}, the dashed red curve is the expected exclusion including all cuts except the one on event-subjettiness, and the solid red curve is the exclusion when $T_{43} < 0.6$ is imposed.  A systematic error $\sys = 20\%$ is assumed for the background prediction.  Cutting on event-subjettiness improves the reach by $\sim350 \mbox{ GeV}$.}
\label{fig: limit4top}
\end{figure}

We begin by considering gluino pair production with $\go \rightarrow t\,\bar{t}+3\,j$.  This topology can yield up to 18 partons when the tops decay hadronically.  For this final state, the $T_{43}$ event-subjettiness variable is most effective.  For a $750 \GeV$ gluino, a cut of $T_{43} < 0.6$ increases $S/B$ from 0.06 to 0.42, and $S/\sqrt{B}$ from 4.9 to 5.6 as seen in Table~\ref{tab: cuts}.  Figure~\ref{fig: limit4top} shows the expected reach for 5 fb$^{-1}$ of 8 TeV data.  The gray line is the NLO gluino pair-production cross section, as evaluated by \texttt{Prospino}.  The dashed red line shows the expected limit when all cuts are applied, except that on event-subjettiness.  With the additional cut on $T_{43}$, the expected limit improves by $\sim 350 \GeV$, as illustrated by the solid red line.  Requiring jets with accidental substructure significantly extends the reach beyond a search that relies on total jet mass alone.            

Event-subjettiness is an effective variable for other RPV gluino decay chains.  However, as the number of hard partons decreases, the signature of accidental substructure becomes more subtle.  Consider the middle diagram of Fig.~\ref{fig:decayDiagrams} where $\go \rightarrow t+2\,j$.  The 8 TeV, 5 fb$^{-1}$ expected limits on this final state are extended from $400 \mbox{ GeV}$ to $600 \mbox{ GeV}$ when $T_{43} < 0.6$ is required in addition to a jet mass cut. For a $500 \mbox{ GeV}$ gluino, cutting on substructure improves the signal to background ratio from 0.14 to 0.61 as seen in Table~\ref{tab: cuts}. Due to the smaller number of partons, the improvement in significance is not as dramatic as for the $\go \rightarrow t\,\bar{t}+3\,j$ topology described previously. Here, the main advantage of cutting on substructure is to increase $S/B$.  This provides a significant improvement because systematic uncertainties tend to drive the sensitivity in the signal region when QCD is the dominant background.
\begin{figure}[b]
\centering   
\includegraphics[scale=0.38]{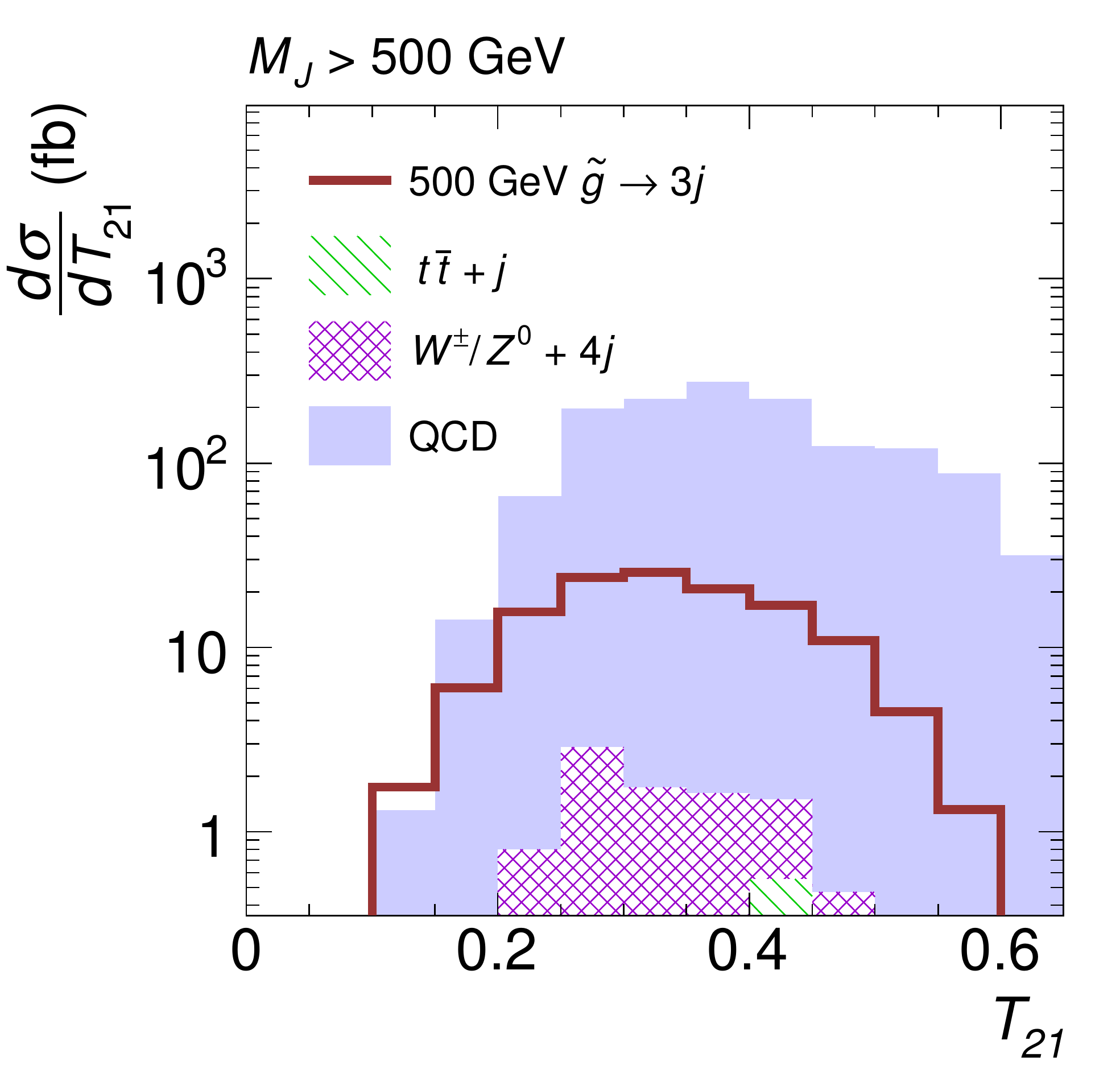} $\quad$ \includegraphics[scale=0.38]{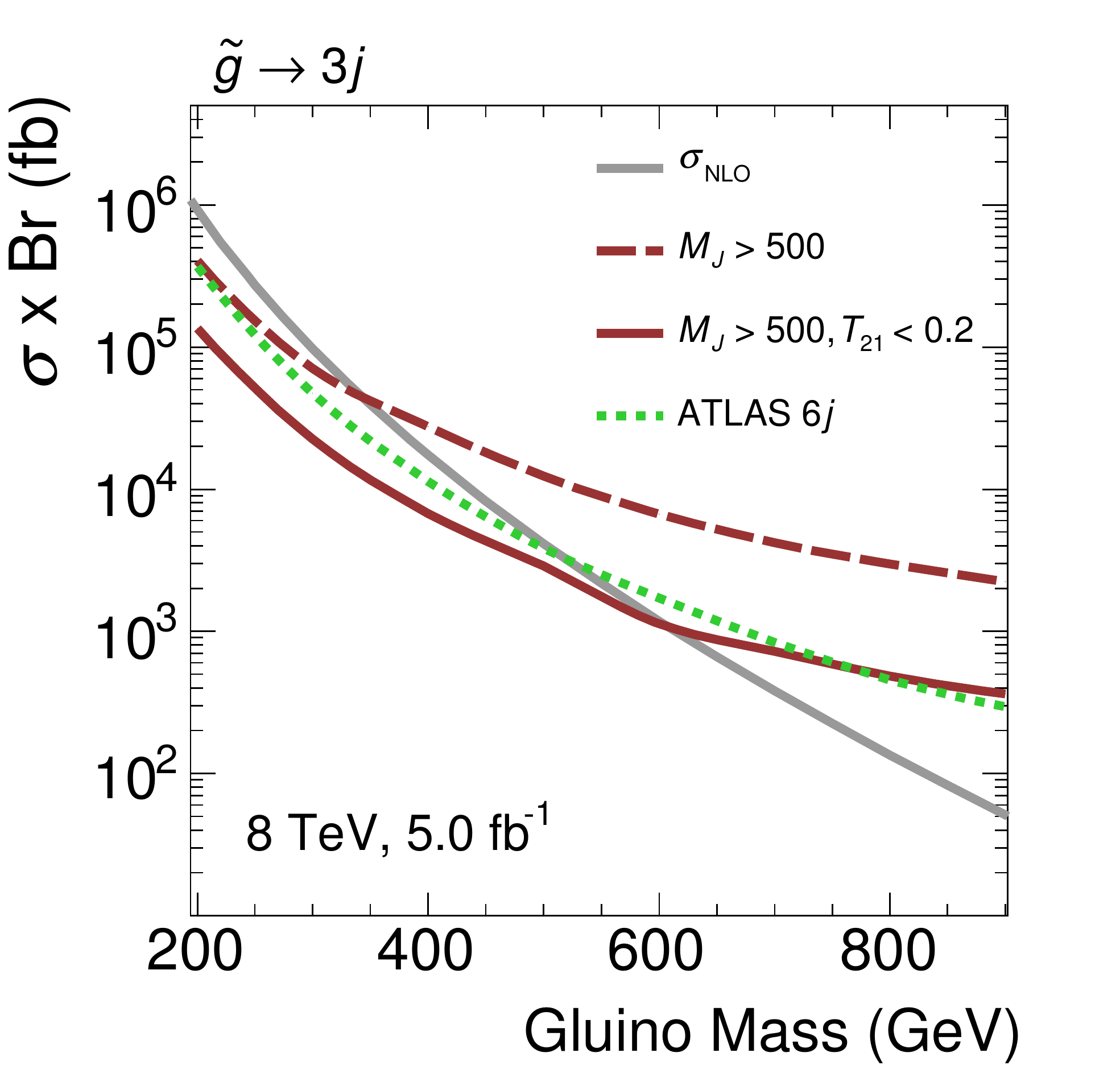}
\caption{The $T_{21}$ distribution for signal and background after requiring $M_J>500 \GeV$ (left) and 95 \% expected exclusion (right) for the $\widetilde{g} \rightarrow 3\,j$ topology at the 8 TeV LHC with 5 fb$^{-1}$ of data.\\
{\bf left:}  The signal (red solid line) is pair production of a $500 \GeV$ gluino with $\widetilde{g} \rightarrow3\,j$.  The stacked histogram is for background (QCD in solid blue, $W^\pm/Z^0 + 4\,j$ in hatched magenta, and $t\,\bar{t}+j$ in striped green).  A cut on $T_{21} \lesssim 0.2$ effectively distinguishes signal from background, after requiring $M_J > 500 \GeV$.\\
{\bf right:}  The solid grey curve is the NLO prediction for the gluino pair production cross section computed using \texttt{Prospino}, the dashed red curve is the expected exclusion including all cuts except the one on event-subjettiness, and the solid red curve is the exclusion when $T_{21} < 0.2$ is imposed.  For comparison, the green dotted line shows our reproduction of the ATLAS search for this same topology~\cite{ATLAS:2012dp}.  Our analysis is competitive with the ATLAS reach.  A systematic error $\sys = 20\%$ is assumed for the background prediction.  A cut on event-subjettiness improves the reach by $\sim 250 \GeV$.  
}
\label{fig: rpvexclusionlimits}
\end{figure}

Lastly, we consider the 6-parton topology illustrated in the right-most diagram of Fig.~\ref{fig:decayDiagrams}.    Of the three decay modes studied in this work, this has the fewest partons and is therefore the most challenging to observe.  In particular, $T_{21}$ provides the best discriminating power for this topology.  The left panel of Fig.~\ref{fig: rpvexclusionlimits} shows the $T_{21}$ distribution for background and signal after applying a $M_J > 500 \GeV$ cut.  The background is peaked between \mbox{$0.35$--$0.4$} and the signal is peaked at \mbox{$0.25$--$0.35$}.  The right panel of Fig.~\ref{fig: rpvexclusionlimits} shows the expected exclusion for the 6-parton final state, assuming $5 \text{ fb}^{-1}$ of 8 TeV data.  The dashed red line shows that the expected limit is $\sim 350$ GeV before a cut on event-subjettiness.  The expected limit increases to $\sim$ 600 GeV when $T_{21} < 0.2$ is required (the solid red line).  As in the last example, the improvement in the limit arises from an increase in the ratio of signal to background after substructure cuts.

The expected reach of our substructure analysis for RPV gluinos is $\sim 600 \GeV$ and compares favorably with published limits from CMS and ATLAS.  The CMS search for three-jet invariant mass resonances~\cite{:2012gw} excludes an RPV gluino from 280--460 GeV with $5 \text{ fb}^{-1}$ of 7 TeV data.  The ATLAS analysis for this final state, published with $4.6 \text{ fb}^{-1}$ of 7 TeV data, uses two techniques to provide exclusions~\cite{ATLAS:2012dp}.  They perform a boosted gluino analysis that makes use of jet substructure and can exclude the gluino in the range 100--255 GeV.\footnote{The recent theory work in~\cite{Curtin:2012rm} finds that the limit on boosted RPV gluinos can be increased by searching for a peak in the jet mass spectrum.}  A separate ``resolved" analysis uses the $p_T$ of the sixth jet (anti-$k_T$, $R=0.4$) to separate signal from background, and excludes the gluino from 100--666 GeV.  

To provide a direct comparison, we reproduce the ATLAS resolved analysis by reclustering our background and signal into anti-$k_T$ jets with $R=0.4$ and applying the cuts from~\cite{ATLAS:2012dp}.  The projected limit for $5 \text{ fb}^{-1}$ of 8 TeV data is shown by the green dotted line of Fig.~\ref{fig: rpvexclusionlimits} and gives a limit of about 550 GeV.\footnote{Note that our expected limit of 550 GeV is weaker than that in~\cite{ATLAS:2012dp}, although it does fall at the edge of the published 1-sigma uncertainty.  We can reproduce their limit if we take a $K$-factor of $1.0$ for the QCD background.  For consistency with the validation plots from Appendix~\ref{sec:Validation}, we use the more conservative 1.8 $K$-factor for Fig.~\ref{fig: rpvexclusionlimits}.}  This demonstrates that our projected limit, which relies on accidental substructure is competitive to that from the ATLAS resolved analysis.

To emphasize the effectiveness of our approach, we also performed a naive comparison between our method and the ATLAS resolved jet analysis of \cite{ATLAS:2012dp} as applied to the $\widetilde{g} \rightarrow t\,\overline{t}+3\,j$ topology.  The ATLAS search is not optimized for this signal; in particular, for this topology relying on $b$-jets and/or leptons may be a more effective strategy.  However, it provides a rough guide for a small-radius jet (with $R\sim0.4$) analysis that one might consider when searching for this multitop topology.  We find that there is no bound on the gluino mass for the 6-jet cuts proposed in \cite{ATLAS:2012dp}.  In principle, the signal region could be extended to a larger jet count. In that case, however, background estimation can be quite challenging. On the other hand, the accidental substructure analysis outlined in this paper is broadly applicable to signals with different jet multiplicities. 

\section{Conclusion}
\label{section:conclusion}

In this paper, we introduced the concept of accidental substructure and illustrated its usefulness in searches for high-multiplicity final states and no missing energy.  Accidental substructure arises because there is a high likelihood that several final-state partons will be clustered together in the same large-radius jet.  These final state partons need not have originated from the same parent particle. QCD is the dominant background.  Having several partons in a QCD event that undergo a large-angle, hard splitting is rare enough to make accidental substructure a useful discriminator.

We analyzed three RPV gluino decay topologies with as many as 18, 10, and 6 partons in the final state.  The requirement that the total jet mass be greater than $500 \GeV$, in conjunction with a cut on event-subjettiness, proved to be very effective.  We found projected limits of $\OO(800 \GeV)$ for the $\widetilde{g} \rightarrow t\,\overline{t}+3\,j$ topology, $\OO(600 \GeV)$ for the $\widetilde{g} \rightarrow t+2\,j$ topology, and $\OO(600 \GeV)$ for the $\widetilde{g} \rightarrow 3\,j$ final state with 5 fb$^{-1}$ of 8 TeV data.  These projections assume a 20\% systematic uncertainty and a conservative $K$-factor for the normalization of the QCD background.  Our goal was to illustrate the general applicability of a search using accidental substructure and we expect that many aspects of this analysis can be further optimized.  One possibility, for instance, is to use a neural network to select the appropriate $N$-subjettiness variables to include in the evaluation of event-subjettiness.  Also, we have not explored how the sensitivity of the search depends on jet radius.  

In the case of the 6-parton final state from RPV gluino decays, our expected limit is comparable to that set by the ATLAS small-radius jet analysis \cite{ATLAS:2012dp}.  Determining the normalization of the QCD background for a $6$ (or more) small-radius jet signal is challenging.  As a result, it is important to have a complementary search with  independent systematics.  Our accidental substructure search is one possible example and is, in addition, sensitive to a broader array of signals than the ATLAS search.  In particular, its sensitivity only improves as the number of final-state partons increases, as we showed for the 10 and 18-parton final states.

Events with many tops can lead to many jets in the final state (the scenario we consider here), but other decay channels can give leptons and \MET$\!\!$. Analyses that tag on a lepton and several $b$-jets can be sensitive in these cases~\cite{Chatrchyan:2012af}. We also expect our reach to improve significantly when $b$-tags are included~\cite{:2012hd}. Alternatively, the total energy $S_T$ may be useful; while it provides the greatest discriminating power in black hole searches~\cite{Chatrchyan:2012taa,Aad:2012ic}, the $S_T$ cut must be above several TeV to adequately reduce the multijet background.  Tagging on a lepton in addition to six or more jets, could allow an $S_T$ cut down to $\sim$~1 TeV~\cite{Lisanti:2011tm}.  

The search we proposed here is complementary to these types of analyses.  We expect that its potential reach will only increase by adding additional handles.  For example, we find that naive cuts on jet mass and event-subjettiness lead to a limit on $\go \rightarrow t\,\bar{t} \,+\!$ \MET that is only slightly weaker than the current bounds from CMS and ATLAS.  Adding a lepton, a $b$-tag and/or a small cut on \MET could make the search even more powerful.  

A significant advantage of using fat jets to study final states with many partons is that it is compatible with data-driven determinations of the QCD background.  Mapping out the phase-space of high multiplicity QCD with Monte Carlo is currently not possible.  For a fat jet analysis, one can use a dijet sample to map out distributions of the internal structure of the jets and to obtain templates for jet mass and substructure as a function of the jet kinematics.  Under the mild assumption that the correlations between fat jets are small, one only needs to predict the phase space distribution of the four fat jets, while the internal properties of each fat jet can be modeled using the template functions derived from the dijet events.  This simple algorithm allows an extrapolation of the QCD contribution to the four fat jet signal region.
    
The possibility of using a jet's internal structure to learn about its origin provides exciting opportunities for new physics searches at the LHC.  Although jet substructure has only been used for boosted signals thus far, this work demonstrates that it is also applicable in the non-boosted regime.  We have shown that accidental substructure provides a robust and powerful new paradigm for new physics searches at the LHC, complementing and extending the reach of current analyses.

\pagebreak

\section*{Note Added}
A related work will appear~\cite{JayToAppear}, which proposes a method of subjet counting and applies it to searches for high-multiplicity signals.

\section*{Acknowledgements}
We thank R. D'Agnolo, D. Curtin, S. El-Hedri, R. Essig, M. Jankowiak, D. Miller, M. Pierini, B. Shuve, and B. Tweedie for useful discussions.  We are especially thankful to S. Hoeche for his help with the QCD Sherpa generation, P. Skands for his help with \mbox{\texttt{Pythia 8}}, and J. Wacker for many conversations on data-driven approaches.  TC is supported by the US Department of Energy under contract number DE-AC02-76SF00515. EI is supported in part by the Government of Canada through Industry Canada, by the Province of Ontario through the Ministry of Research and Information, by the US Department of Energy under contract number DE-AC02-76SF00515, and by an LHCTI graduate fellowship under grant NSF-PHY-0969510. ML is supported by the Simons Postdoctoral Fellows Program.   HKL is supported by the DOE Office of Science Graduate Fellowship Program.  TC and ML would like to thank the Aspen Center for Physics, which is supported by the National Science Foundation grant number 1066293.  ML also acknowledges the Galileo Galilei Institute for Theoretical Physics for their hospitality.
\pagebreak

\appendix
\section{Simulation Details and Validation}\label{sec:Validation}
In this appendix, we discuss the details of our simple detector mockup and provide validation plots comparing our QCD  Monte Carlo to a number of public distributions from ATLAS.  We extract a $K$-factor to normalize our QCD sample and demonstrate that our Monte Carlo reproduces the measured shapes of substructure and jet mass distributions to reasonable accuracy. 

We simulate detector granularity by clustering stable, visible generator-level particles into $\eta\times \phi$ cells of size $0.1\times 0.1$.  Electrons, muons, and photons are kept if they fall within $|\eta| < 2.5$, while all other particles are kept if they fall within $|\eta| < 3.0$.  Each calorimeter cell is assigned a light-like vector with energy equal to the sum of all particle energies contained therein.  \texttt{FastJet 3.0} clusters these four-vectors into anti-$k_{T}$ jets and computes $N$-subjettiness for the resulting jets using the ``min\_axes" algorithm, implemented in the $N$-subjettiness plugin of Thaler and Van-Tilberg~\cite{Thaler:2010tr,Thaler:2011gf}.  Note that leptons are included in jet clustering and when calculating substructure variables.  A jet is removed if it is within $\Delta R<0.2$ of a lepton and its $p_{T}$ is less than twice the lepton's $p_T$.  

We validate our QCD Monte Carlo by comparing against published kinematic and substructure distributions.  No published 8 TeV substructure results are currently available, and so we compare against the published 7 TeV ATLAS results~\cite{ATLAS:2012am,ATLAS:2012dp,Aad:2011tqa}.  A weighted sample of $p\,p\rightarrow n\,j$, where $n\in (2,\dots,6)$, is generated in \texttt{Sherpa 1.4.0}.  Our 7 TeV sample consists of $\sim 50$ million events and is generated with the same settings as our $\sim 400$ million event 8 TeV Sherpa sample, described in Sec.~\ref{section: MonteCarlo}.

\begin{figure}[b]
\centering
\includegraphics[scale=0.37]{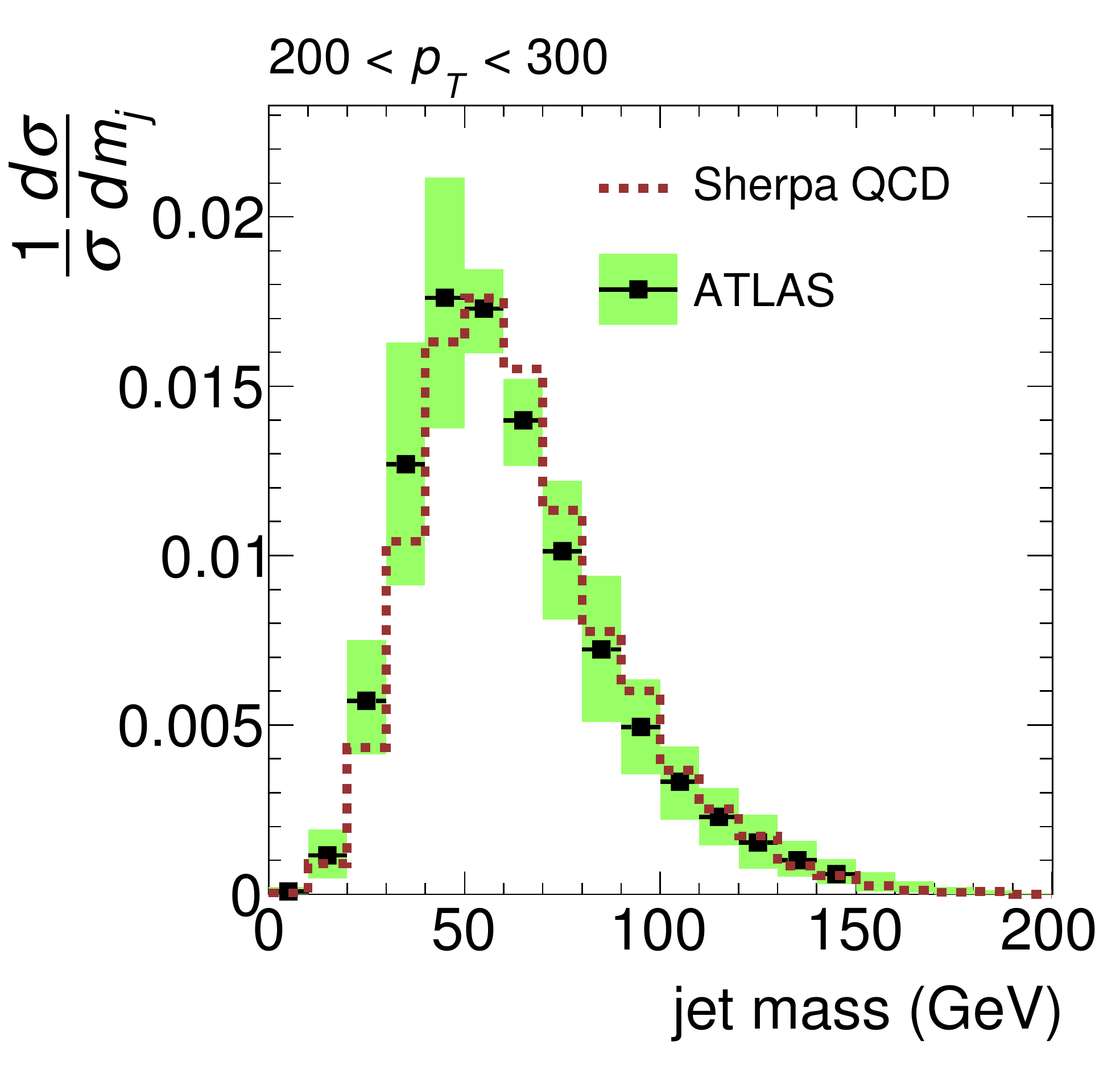}\\
\includegraphics[scale=0.37]{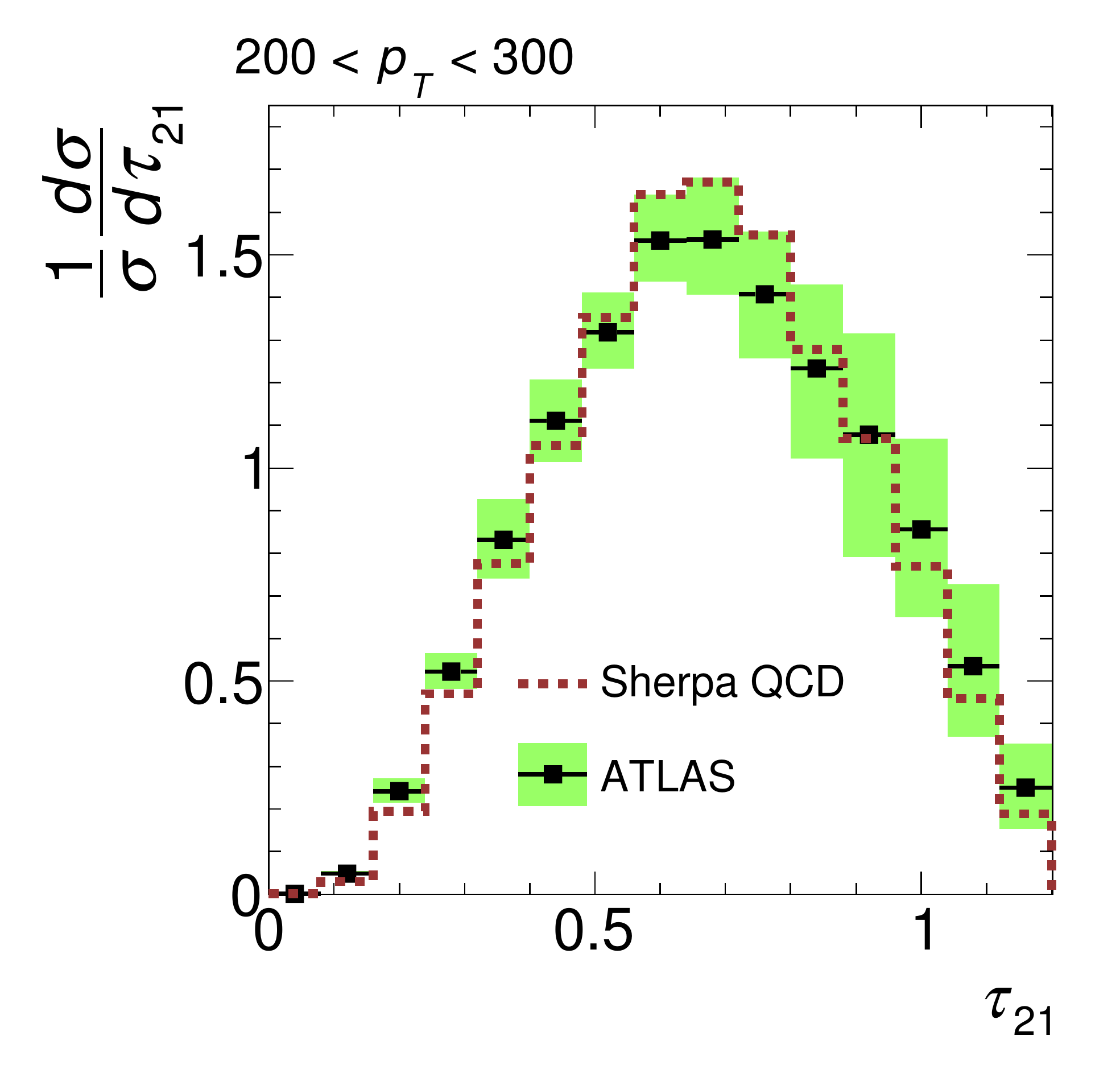}
\includegraphics[scale=0.37]{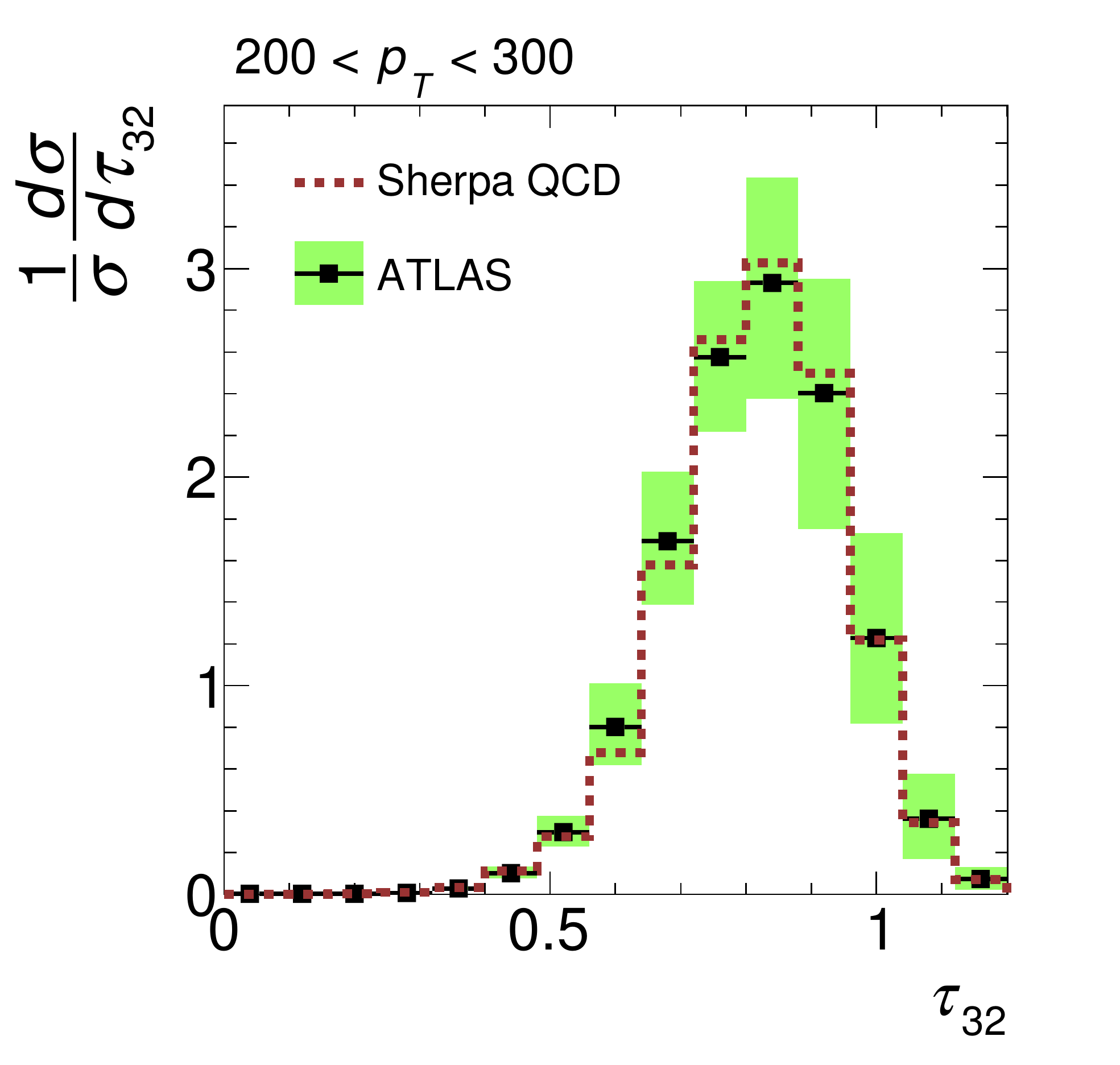}
\caption{Jet mass (top) and $N$-subjettiness (bottom) comparisons between the \texttt{Sherpa} QCD prediction [dotted red] and the ATLAS results [black rectangle] of~\cite{ATLAS:2012am}.   The green band is the combined statistical and systematic error in the ATLAS measurement including the uncertainty from the unfolding procedure.}
\label{fig:ValidateJetMassAndNSubj}
\end{figure}

To validate the shape of the jet mass and substructure distributions, we follow the analysis in~\cite{ATLAS:2012am} and compare to the unfolded distributions.  Particles are clustered into anti-$k_{T}$ jets with $R=1.0$.  The resulting jets are divided into four equally-spaced $p_{T}$ bins from 200 to 600 GeV.  The jet mass ($\tau_{21}$ and $\tau_{32}$) distributions are shown in the top (bottom) of Fig.~\ref{fig:ValidateJetMassAndNSubj} for $p_{T} \in (200,300)$.  The Monte Carlo predictions are well within the error bands quoted by ATLAS.  We checked that the \texttt{Sherpa} results for the higher $p_{T}$ bins, not shown here, also match the ATLAS results.  

\texttt{Sherpa} outputs a leading order (matched) cross section of $\sigma_\text{QCD}^\text{Sherpa} = 9.6\times10^9 \text{ fb}$.  Because this cross section is enhanced by loop effects, we must find the proper normalization, or $K$-factor, for the QCD background: 
\begin{equation}
\sigma_\text{QCD} = K\times \sigma_\text{QCD}^\text{Sherpa}.
\end{equation}
Using the reported 2-jet inclusive cross-section in~\cite{Aad:2011tqa}, we obtain a $K$-factor of $\sim 1.3$.  Comparing to the $6^{\textrm{th}}$ jet $p_{T}$ distribution in~\cite{ATLAS:2012dp}, we obtain a $K$-factor of $1.8$.  Furthermore, by comparing the normalization of the jet mass, $\tau_{21}$ and $\tau_{32}$ distributions in \cite{ATLAS:2012am} we obtain a $K$-factor of $1.8$.  To be conservative, we assume a $K$-factor of $1.8$ in this work.

\pagebreak

\bibliography{substructure}
\bibliographystyle{utphys}

\end{document}